\documentclass[aps,prl,twocolumn,notitlepage]{revtex4-1}

\usepackage{amsmath}
\usepackage{amssymb}
\usepackage{graphicx}
\usepackage{epstopdf}
\DeclareMathOperator{\sgn}{sgn}
\usepackage[colorlinks=true]{hyperref}

\begin{document}
\title{Chiral edge mode in the coupled dynamics of magnetic solitons in a honeycomb lattice}

\author{Se Kwon Kim}
\author{Yaroslav Tserkovnyak}
\affiliation{Department of Physics and Astronomy, University of California, Los Angeles, California 90095, USA}

\date{\today}

\begin{abstract}
Motivated by a recent experimental demonstration of a chiral edge mode in an array of spinning gyroscopes, we theoretically study the coupled gyration modes of topological magnetic solitons, vortices and magnetic bubbles, arranged as a honeycomb lattice. The soliton lattice under suitable conditions is shown to support a chiral edge mode like its mechanical analogue, the existence of which can be understood by mapping the system to the Haldane model for an electronic system. The direction of the chiral edge mode is associated with the topological charge of the constituent solitons, which can be manipulated by an external field or by an electric-current pulse. The direction can also be controlled by distorting the honeycomb lattice. Our results indicate that the lattices of magnetic solitons can serve as reprogrammable topological metamaterials.
\end{abstract}

\maketitle

\emph{Introduction.}|The term metamaterials refer to a class of man-made composite materials which can offer functionalities beyond those found in nature via collective dynamics of constituent elements~\cite{*[][{, and references therein.}] MaSA2016}. Inspired by the robust edge states in the topological electronic phases such as quantum Hall states~\cite{*[][{, and references therein.}] Prange1990}, topological metameterials with analogous edge states have been proposed and realized in optical~\cite{HaldanePRL2008, *LuNP2014}, acoustic~\cite{MaSA2016}, magnetic~\cite{ShindouPRB2013, *ShindouPRB2013-2}, and mechanical systems~\cite{ProdanPRL2009, *ZhangPRL2010, *SunPNAS2012, *KaneNP2014, *ChenPNAS2014, *PauloseNP2015, *SusstrunkScience2015}. In particular, it has recently been shown theoretically~\cite{WangPRL2015} and experimentally~\cite{NashPNAS2015} that a honeycomb lattice of spinning gyroscopes can support a chiral edge mode that is protected from small perturbations such as lattice distortions and thus can be identified as a topological mechanical metamaterial. As discussed in Ref.~\cite{NashPNAS2015}, an open challenge for its practical applications is to find a feasible way to keep gyroscopes spinning.

Quantum-mechanically, nature has already endowed us a permanent gyroscope: spin of a particle. This intrinsic angular momentum manifests itself macroscopically through the gyrotropic force in the dynamics of magnetic solitons with topologically nontrivial textures such as magnetic bubbles (also known as skyrmions) and vortices~\cite{ThielePRL1973}. These solitons and their dynamics have attracted much attention of physicists due to their fundamental properties~\cite{*[][{, and references therein.}] KosevichPR1990} and technological promise~\cite{*[][{, and references therein.}] BaderRMP2006, *[][{, and references therein.}] NagaosaNN2013}. In particular, the collective gyration modes of arrays of vortex disks have been studied theoretically~\cite{ShibataPRB2003, *ShibataPRB2004, *GalkinPRB2006, *LeeJAP2011, *SukhostavetsPRB2013} and experimentally~\cite{BarmanJPD2010, *BarmanIEEE2010, *VogelPRL2010, *JungSR2011, *StreubelPRB2015, *AdolffPRB2015, SugimotoPRL2011, HanSR2013, BehnckePRB2015} as reprogrammable metamaterials whose functionalities can be controlled by changing vortices' polarities and chiralities~\cite{TaniuchiJAP2005, *ChoiAPL2007, *ChoiAPL2010}.

When viewing topological magnetic solitons as gyroscopes, it is natural to expect that a honeycomb lattice of the solitons can support a chiral edge mode as its mechanical analogues~\cite{NashPNAS2015, WangPRL2015}. In this Letter, we verify the expectation both by numerically solving the equations of motion for the dynamics of coupled solitons and by mapping the system to the Haldane model for an electron in graphene, which is known to exhibit the quantum Hall effect~\cite{HaldanePRL1988-2}. We also show that the direction of the edge mode can be controlled either by changing the topological charge of the solitons or by distorting the geometry of the honeycomb lattice. We conclude the Letter with an experimental outlook, including a possibility of the thermal chirality control using ferrimagnets~\cite{KirilyukRMP2010, *KirilyukRPP2013}.

\begin{figure}
\includegraphics[width=\columnwidth]{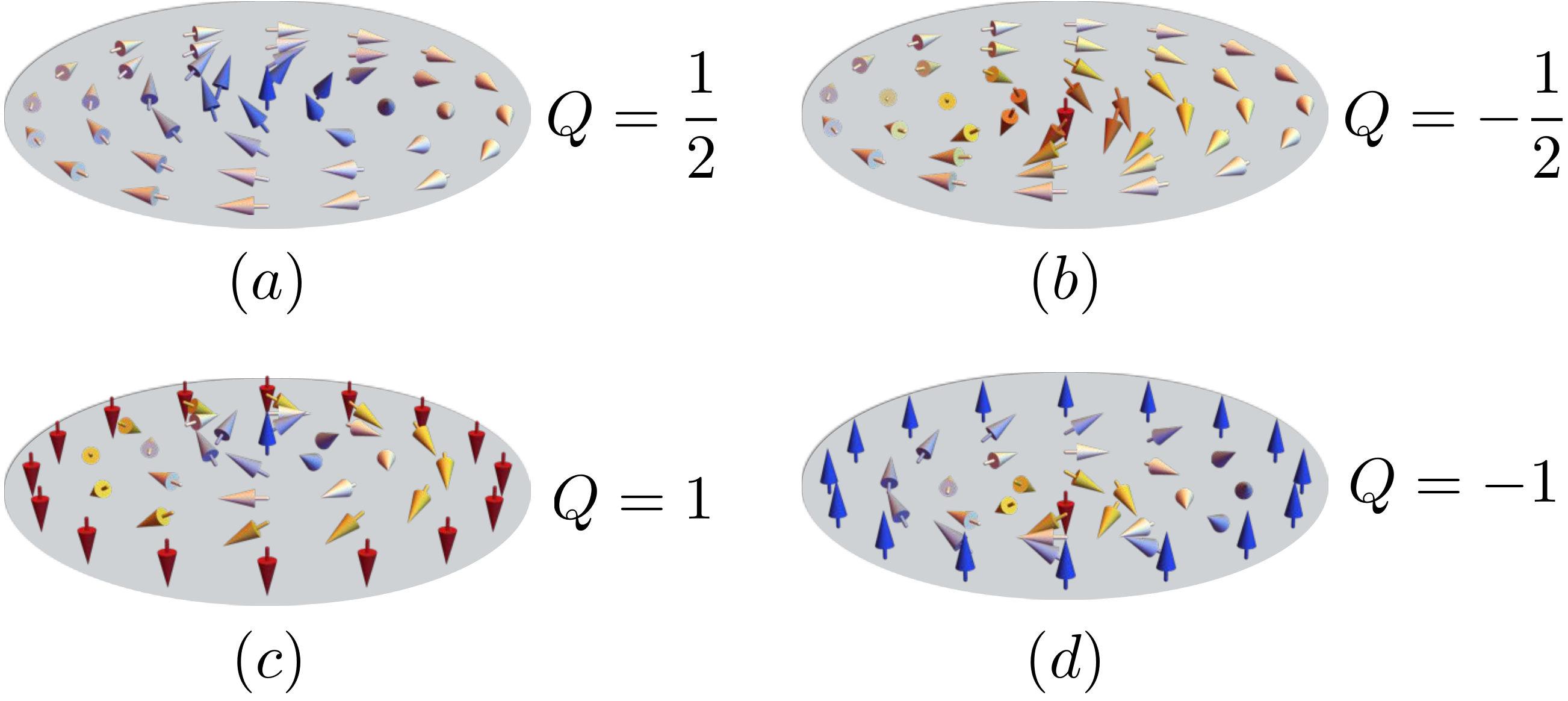}
\caption{Schematic illustrations of (a) a vortex with the topological charge $Q = 1/2$, (b) a vortex with $Q = -1/2$, (c) a magnetic bubble with $Q = 1$, and (d) a magnetic bubble with $Q = -1$.}
\label{fig:fig1}
\end{figure}

\emph{Model.}|We consider a two-dimensional array of magnetic solitons such as vortices and magnetic bubbles, which are characterized by their topological charges,
\begin{equation}
Q \equiv \frac{1}{4 \pi} \int dx dy \, \mathbf{n} \cdot (\partial_x \mathbf{n} \times \partial_y \mathbf{n}) \, ,
\end{equation}
which measures how many times the unit vector $\mathbf{n}$ along the direction of the local magnetization wraps the unit sphere. The elementary topological charges of vortices and magnetic bubbles are $Q = \pm 1/2$ and $Q = \pm 1$, respectively, the sign of which is determined by the internal structure. See Fig.~\ref{fig:fig1} for schematic illustrations of them. The slow motion of the solitons can be described by their positions, $\mathbf{R}_j \equiv (X_j, Y_j)$, which are assumed to be subjected to the restoring force toward equilibrium positions, $\mathbf{R}^0_j \equiv (X^0_j, Y^0_j)$. The low-energy dynamics of the coupled solitons can be described by Thiele's equation~\cite{ThielePRL1973} within the approximation of the rigid soliton texture:
\begin{equation}
\label{eq:eom}
G \hat{\mathbf{z}} \times \dot{\mathbf{U}}_j - \alpha D \dot{\mathbf{U}}_j + \mathbf{F}_j = 0 \, ,
\end{equation}
where $\mathbf{U}_j \equiv \mathbf{R}_j - \mathbf{R}^0_j$ is the displacement of the soliton from the equilibrium position, $G \equiv - 4 \pi s t Q$ is the gyrotropic coefficient, $s \equiv M_s / \gamma$ is the spin density of the magnet, $M_s$ is the saturation magnetization, $\gamma$ is the gyromagnetic ratio, $t$ is the thickness of the magnet, $\alpha D \equiv \alpha c s t$ is the viscous coefficient, $c$ is a dimensionless geometric factor determined by the exact profile of the solitons, $\alpha$ is the Gilbert damping constant~\cite{GilbertIEEE2004}, and $\mathbf{F}_j \equiv - \partial U / \partial \mathbf{U}_j$ with $U$ the potential energy as a function of the displacements. Here, the first term is the gyrotropic force proportional the spin density, which is crucial for the analogy between the lattice of magnetic solitons and the lattice of mechanical gyroscopes; the second term is the viscous force; the third term is the conservative force.

\begin{figure}
\includegraphics[width=\columnwidth]{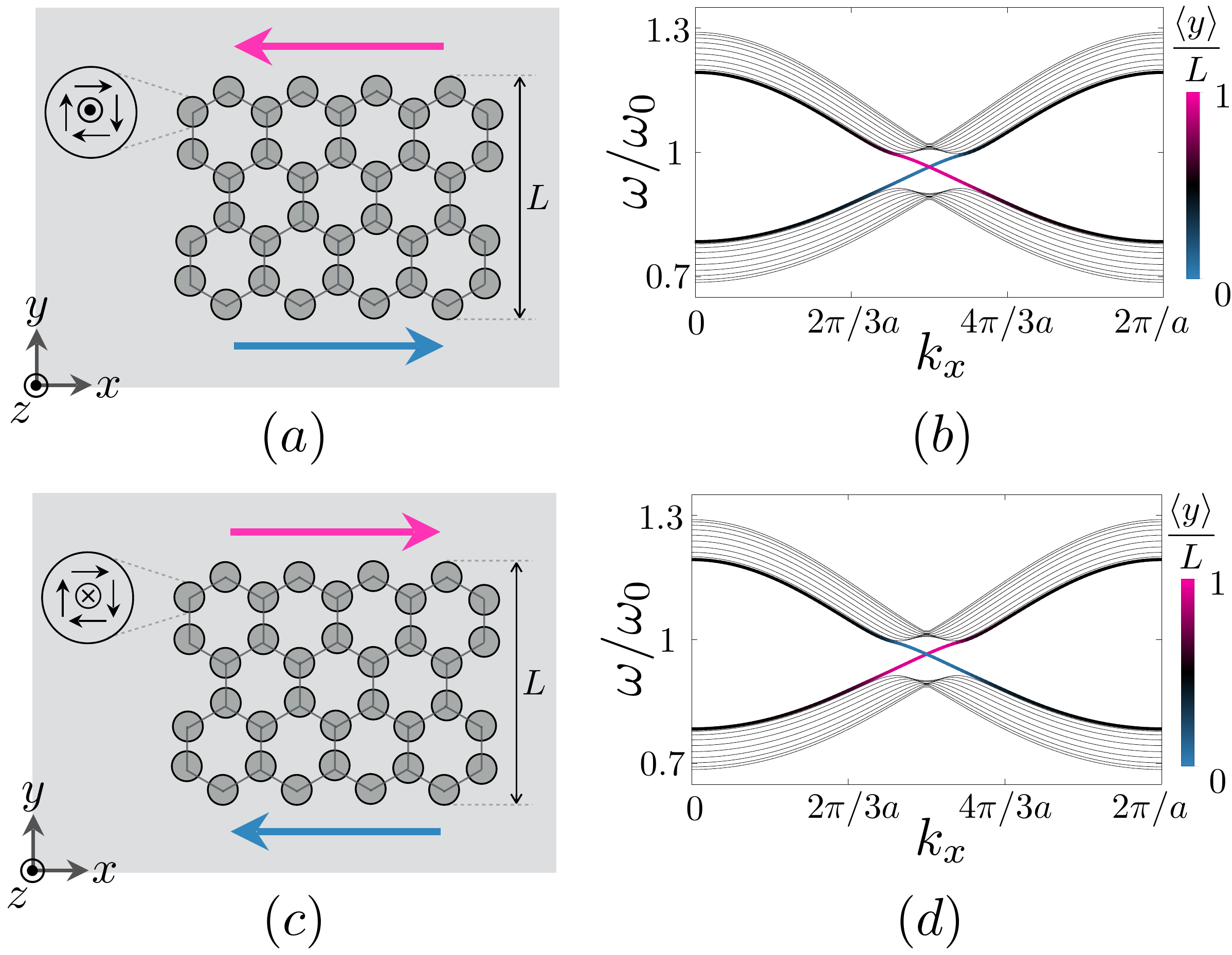}
\caption{(a) A schematic illustration of physically separated ferromagnetic disks (drawn as circles) in a honeycomb lattice with zigzag edges; vortices in disks have the polarity $p = 1$ and thus the topological charge $Q = 1/2$. The gray lines between circles represent the magnetostatic interactions between vortices. The arrows along the edges represent the directions of the chiral mode. (b) The one-dimensional dispersion for the coupled gyration modes of the system shown in (a), which is obtained by solving Eq.~(\ref{eq:eom2}) numerically. The symbol $a$ represents the distance between the second-nearest neighbors. The color represents the average vertical position $\langle y \rangle$ weighted by the amplitude squared corresponding to each mode. (c), (d) Analogous figures for vortices with the polarity $p = -1$, corresponding to the topological charge $Q = -1/2$.}
\label{fig:fig2}
\end{figure}

To the quadratic order in the displacements, the energy of the system is modeled by $U = \sum_j K \mathbf{U}_j^2 / 2 + \sum_{j \neq k} U_{jk} / 2$, where the first term is the pinning potential parametrized by the spring constant $K > 0$ and the second term is the interaction between two solitons. The following form is considered for the interactions:
\begin{equation}
\label{eq:U}
U_{jk} = I_{\parallel} (d_{jk}) U_j^\parallel U_k^\parallel - I_{\perp} (d_{jk}) U_j^\perp U_k^\perp \, .
\end{equation}
Here, $d_{jk} \equiv |\mathbf{R}_j^0 - \mathbf{R}_k^0|$ is the distance between two solitons in the absence of the interaction; $u_j^\parallel \equiv \hat{\mathbf{e}}_{jk} \cdot \mathbf{U}_j$ is the projection of the displacement $\mathbf{U}_j$ onto the line connecting two solitons, described by the unit vector $\hat{\mathbf{e}}_{jk} \equiv (\mathbf{R}_k^0 - \mathbf{R}_j^0) / d_{jk}$; $U_j^\perp \equiv (\hat{\mathbf{z}} \times \hat{\mathbf{e}}_{jk}) \cdot \mathbf{U}_j$ is the projection of $\mathbf{U}_j$ onto the line perpendicular to $\hat{\mathbf{e}}_{jk}$; $I_\parallel (d_{jk})$ and $I_\perp (d_{jk})$ parametrize the corresponding interactions, which are attractive (repulsive) if the value is positive (negative). We will simplify the subsequent discussion by assuming that the interactions are much weaker than the pinning potential, $| I_\parallel |, | I_\perp | \ll K$. The interaction of this form can capture the magnetostatic interactions between two vortices in separated disks~\cite{ShibataPRB2003} and the exchange-mediated interactions between two magnetic bubbles, as will be explained further below. Thiele's equation (\ref{eq:eom}) with the inter-vortex interactions in Eq.~(\ref{eq:U}) has been employed successfully to describe the observed dynamics of an array of vortex disks~\cite{BarmanJPD2010, SugimotoPRL2011}.

We are interested in the dispersion of the normal modes of the coupled gyration dynamics of solitons. Since the main effects of the viscous force with the small Gilbert damping $\alpha \ll 1$ is the broadening of the dispersion linewidth, not the change of the dispersion itself, we will neglect it henceforth. In addition, we will take account of the interactions between nearest neighbors alone by assuming that the interactions decay sufficiently fast as a function of the distance, which will be justified below individually for each physical realization. Then, with the energy given above, Thiele's equations of motion for $\mathbf{U}_j \equiv (u_j, v_j)$ [Eq.~(\ref{eq:eom})] can be written as the following coupled equations for $u_j$ and $v_j$:
\begin{equation}
\label{eq:eom2}
\begin{split}
0 = & \sgn(Q) \begin{pmatrix} \dot{v}_j \\ - \dot{u}_j \end{pmatrix} - \omega_0 \begin{pmatrix} u_j \\ v_j \end{pmatrix} \\
& - \sum_{k \in \langle j \rangle} \begin{pmatrix} \zeta + \xi \cos 2 \theta_{jk} & \xi \sin 2 \theta_{jk} \\ \xi \sin 2 \theta_{jk} & \zeta - \xi \cos 2 \theta_{jk} \end{pmatrix} \begin{pmatrix} u_k \\ v_k \end{pmatrix} \, ,
\end{split}
\end{equation}
where $\omega_0 \equiv K / |G|$ is the gyration frequency of an isolated soliton, $\langle j \rangle$ represents the set of the nearest neighbors of the soliton $j$, $\theta_{jk}$ is the angle of the direction $\hat{\mathbf{e}}_{jk}$ from the $x$ axis, $\zeta \equiv (I_\parallel - I_\perp) / |G|$, and $\xi \equiv (I_\parallel + I_\perp) / |G|$. The assumption that the interaction is much weaker than the pinning potential translates into $\zeta, \xi \ll \omega_0$.

\emph{Vortex honeycomb lattice.}|A ferromagnetic disk of a suitable size can harbor a magnetic vortex in its ground state~\cite{CowburnPRL1999, *ChienPT2007}. A vortex is characterized by the polarity $p = \pm 1$, which is the direction of the magnetization at its core, and the chirality, $c = \pm 1$, which describes the clockwise ($c = -1$) or counterclockwise ($c = 1$) in-plane curling of the magnetization around the core. The polarity of a vortex is related to its topological charge by $Q = p/2$. The polarization and the chirality can be independently controlled by an external field or an electric-current pulse~\cite{TaniuchiJAP2005}. Throughout the Letter, we use vortices with the positive chirality $c = 1$, which are shown in Figs.~\ref{fig:fig1}(a) and (b).

Let us consider vortices in physically separated disks that are arranged as a honeycomb lattice. See Fig.~\ref{fig:fig2} for illustrations. The displacements of vortices from the centers of disks generate the magnetostatic charges by altering the magnetization profile from the ground state~\cite{MetlovJMMM2002, *GuslienkoPRL2006, *MetlovPRL2010}. The magnetostatic energy associated with these charges engender the inter-vortex interaction given by Eq.~(\ref{eq:U})~\cite{ShibataPRB2003}. The magnitudes of the interactions decay as $I_\parallel (d), I_\perp (d) \sim d^{-6}$, which justifies our nearest-neighbor model in Eq.~(\ref{eq:eom2}). Let us take the experimental values for the parameters from Ref.~\cite{SugimotoPRL2011} to obtain the normal-mode dispersion from Eq.~(\ref{eq:eom2}). For two permalloy disks of the radius $R = 500$nm, the thickness $t = 50$nm, and the center-to-center distance $d = 1075$nm, the parameters are given by $K \sim 5 \times 10^{-4}$J/m$^2$, $\omega_0 \sim 2$ GHz, $I_\parallel \sim 5 \times 10^{-5}$J/m$^2$, and $I_\perp ~\sim 3 \times 10^{-5}$J/m$^2$. These measured values agree with the theoretical estimations~\cite{ShibataPRB2003}. The corresponding parameters in our model [Eq.~(\ref{eq:eom2})] are given by $\zeta \sim 0.05 \, \omega_0$ and $\xi \sim 0.15 \, \omega_0$.

We solve Eq.~(\ref{eq:eom2}) for a honeycomb ribbon with periodic boundary conditions along the $x$ direction and with zigzag terminations at the top $y = L$ and the bottom $y = 0$. The color represents the average vertical position of the mode, $\langle y \rangle \equiv \sum_j Y_j^0 |\mathbf{U}_j|^2 / \sum_j |\mathbf{U}_j|^2$. The one-dimensional dispersions for the normal modes are shown in Fig.~\ref{fig:fig2}(b) and (d) for vortices with the polarity $p = 1$ and $p = -1$, respectively. The results show that each system supports the chiral edge mode lying within the bulk gap, which rotates the boundary in the same direction as individual solitons precess. Two polarities, $p = 1$ and $p = -1$, are related by the magnetization flip and thus by the time reversal. The chiralities of the edge modes are opposite accordingly.

\emph{Mapping to the Haldane model.}|The existence of the chiral edge mode can be understood analytically by mapping Eq.~(\ref{eq:eom2}) to the Haldane model of the quantum Hall effect~\cite{HaldanePRL1988-2}. The analogous mapping is given in Ref.~\cite{NashPNAS2015} for mechanical gyroscopes, which we adopt here for magnetic solitons. For simplicity, we will consider the case of the negative topological charge, $Q = -1/2$. We begin by casting Eq.~(\ref{eq:eom2}) in terms of the complex variable, $\psi_j \equiv u_j + i v_j$:
\begin{equation}
i \dot{\psi}_j = \omega_0 \psi_j + \sum_{k \in \langle j \rangle} ( \zeta \psi_k + \xi e^{2i \theta_{jk}} \psi_k^* ) \, .
\end{equation}
The equation can be interpreted as the Schr{\"o}dinger equation for the wavefunction of an electron in a tight-binding model. Since the interactions are much weaker than the pinning potential, $\zeta, \xi \ll \omega_0$, we can use perturbation theory to find the equation in terms of $\psi_j$ alone by eliminating its complex conjugate. To that end, let us expand the complex variable as $\psi_j (t) = \chi_j (t) \exp(- i \omega_0 t) + \phi_j (t) \exp(i \omega_0 t)$ where $\chi_j(t)$ and $\phi_j(t)$ change over time slowly on the time scale set by the frequency $\omega_0$. Then, by matching the coefficients of $\exp(i \omega_0 t)$ and $\exp(- i \omega_0 t)$ in Eq.~(\ref{eq:eom2}), we obtain 
$
i \dot{\chi}_j = \sum_{k \in \langle j \rangle} (\zeta \chi_k + \xi e^{2i \theta_{jk}} \phi_k^* )
$ and 
$
i \dot{\phi}_j = 2 \omega_0 \phi_j + \sum_{k \in \langle j \rangle} ( \zeta \phi_k + \xi e^{2i \theta_{jk}} \chi_k^* )
$.
Since the interactions are weak, $\zeta, \xi \ll \omega_0$, the soliton dynamics is mostly associated with the resonance frequency $\omega_0$ and thus $|\chi_j| \gg |\phi_j|$. Using the approximation $2 \omega_0 \phi_j \approx - \sum_{k \in \langle j \rangle} \xi e^{2i \theta_{jk}} \chi_k^*$ obtained from the latter yields the following equation:
\begin{equation}
\label{eq:Haldane}
\begin{split}
i \dot{\psi}_j = & \left( \omega_0 - 3 \xi^2 / 2 \omega_0 \right) \psi_j + \zeta \sum_{k \in \langle j \rangle} \psi_k \\
& - (\xi^2 / 2 \omega_0) \sum_{l \in \langle\langle j \rangle\rangle} \cos \left( 2 \bar{\theta}_{jl} \right) \psi_l \\
& - i (\xi^2 / 2 \omega_0) \sum_{l \in \langle\langle j \rangle\rangle} \sin \left( 2 \bar{\theta}_{jl} \right) \psi_l \, ,
\end{split}
\end{equation}
where $\langle\langle j \rangle\rangle$ is the set of the second-nearest neighbors of $j$, $\bar{\theta}_{jl} \equiv \theta_{jk} - \theta_{kl}$ is the relative angle from the bond $k \rightarrow l$ to the bond $j \rightarrow k$ with $k$ between $j$ and $l$. This equation is similar to the Haldane model for an electron in a honeycomb lattice~\cite{HaldanePRL1988-2, KanePRL2005}, which is a prototypical example exhibiting the quantum Hall effect. The difference is the term in the second line, which is real and does not affect the existence of the chiral edge mode~\cite{NashPNAS2015}.

Let us explain how the chiral edge mode originates in the above equation for electrons. The angle between the neighboring bonds is given by $\bar{\theta}_{jl} = \pm 2 \pi / 3$, where the upper (lower) sign is for the case when we have to turn right (left) to go from $j$ to $l$. When the last term vanishes, two electronic bands associated with two sublattices touch each other at two points in the momentum space, forming two Dirac cones. When the last term is finite, an electron picks up a phase when hopping to its second-nearest neighbors and the sign of the accumulated phase depends on whether the electron makes a left or right turn to arrive at its neighbors. Via this path-dependent phase, the imaginary second-nearest hopping opens topological gaps at the Dirac cones and engenders the chiral edge mode. The gap size is given by $\Delta =  9 \xi^2 / 2 \omega_0$~\cite{KanePRL2005}.

\begin{figure}
\includegraphics[width=\columnwidth]{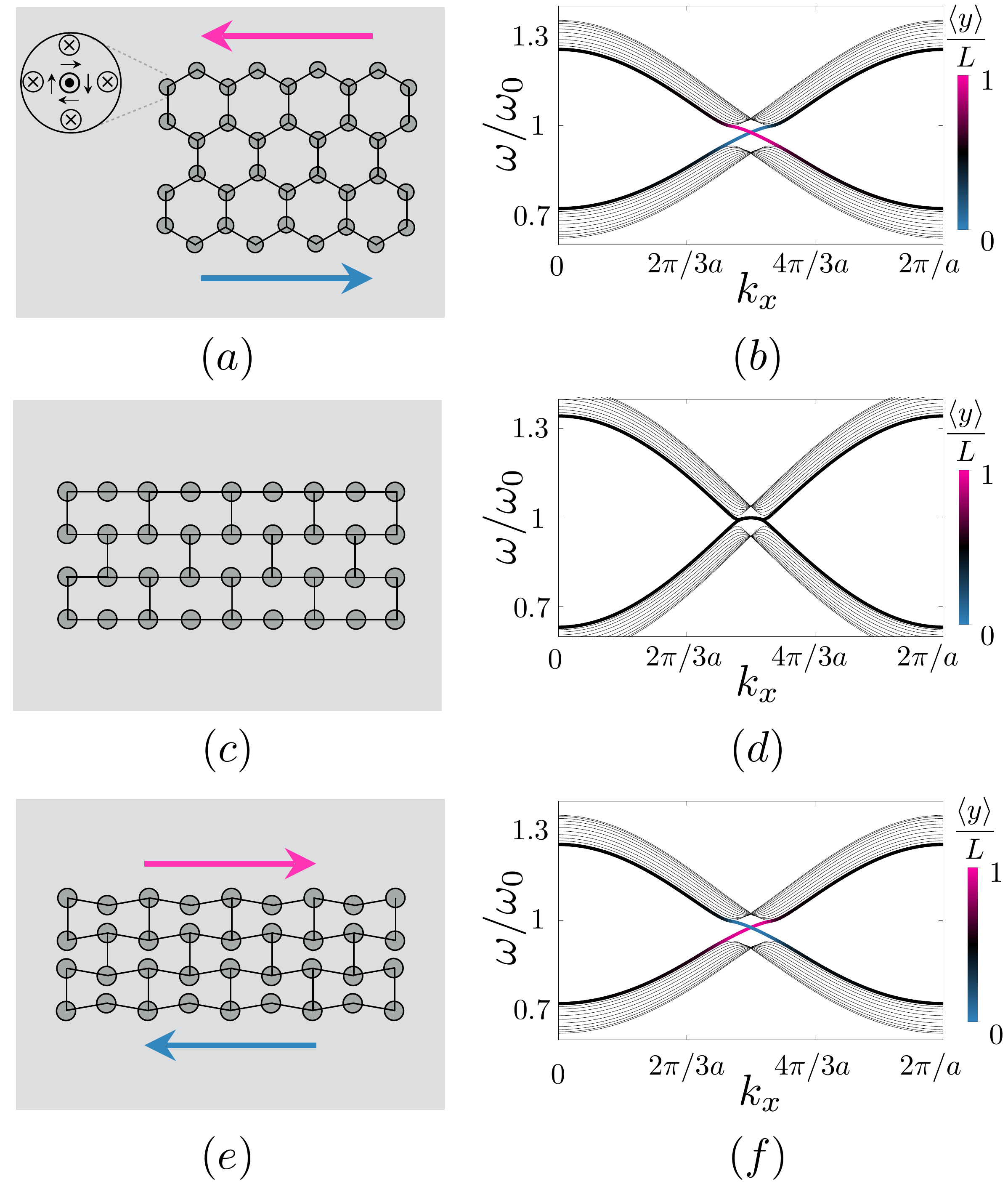}
\caption{(a) A schematic illustration of magnetic bubbles with the topological charge $Q = 1$ (drawn as circles) in a honeycomb lattice with zigzag edges. The black lines between circles represent the exchange-coupled interactions between magnetic bubbles connected by the magnetic strips. (b) The one-dimensional dispersion for the coupled gyration modes for the system shown in (a). (c), (d) Analogous figures when the angles between nearest bonds are multiples of $\pi / 2$, for which the dispersion for the bulk is gapless and thus the topological edge mode is not supported. (e), (f) Analogous figures for the system exhibiting the chiral edge mode in the opposite direction to (a) and (b).}
\label{fig:fig3}
\end{figure}

\emph{Magnetic bubble honeycomb lattice.}|Now let us turn to a honeycomb soliton lattice, in which only the nearest-neighbor solitons connected by bonds are engineered to interact. We consider distortions of the constituent hexagons, which preserve the lattice connectivity and the bond lengths corresponding to the original nearest neighbors. From the above discussions, the origin of the chiral edge mode is the staggered phase gathered by an electron hopping between the second-nearest neighbors~\cite{HaldanePRL1988-2}. As can be seen in Eq.~(\ref{eq:Haldane}), the sign of the phase can be controlled by changing the angle $\bar{\theta}_{jl}$ between the neighboring bonds. Let us take examples shown in Fig.~\ref{fig:fig3}. In Fig.~\ref{fig:fig3}(a), $\sin 2 \bar{\theta}_{jl}$ is positive (negative) if we make a right (left) turn to go from $j$ to $l$; In Fig.~\ref{fig:fig3}(c), $\sin 2 \bar{\theta}_{jl}$ vanishes; In Fig.~\ref{fig:fig3}(e), the sign of $\sin 2 \bar{\theta}_{jl}$ is opposite to the case in Fig.~\ref{fig:fig3}(a) for all pairs of $j$ and $k$. Since the staggered phase changes its sign between (a) and (c) with vanishing in (b), we expect the change of the chirality of the topological edge mode from (a) to (c) via the gap closing in (b)~\cite{NashPNAS2015}. We verify this expected dependence of the chirality on the shape of the constituent hexagons by numerically solving Eq.~(\ref{eq:eom2}) below by taking the approach used in Ref.~\cite{NashPNAS2015}, which has studied the analogous problem for gyroscope lattices.

Let us consider a honeycomb lattice of magnetic bubbles with the topological charge $Q = 1$, which can appear as a ground state of a magnetic disk with perpendicular anisotropy~\cite{Bobeck1975, *Malozemoff1979, *Eschenfelder1981, *HehnScience1996, *FukumuraScience1999}. The constituent disks are connected by ferromagnetic strips so that neighboring magnetic bubbles can interact with each other via the exchange energy. The coupled gyration modes of magnetic bubbles in one-dimensional magnetic strip have been studied by micromagnetic simulations in Ref.~\cite{KimSR2017}, according to which the dominant contributions to the interaction comes from the exchange energy. We model the exchange-driven (repulsive) interaction as a function of the distance, $f(\mathbf{R}_j, \mathbf{R}_k) = f(|\mathbf{R}_j - \mathbf{R}_k|)$ by following Refs.~\cite{LinPRB2013, *PinnaarXiv2017}. To the second order in the displacements, the interaction can be written in the form of Eq.~(\ref{eq:U}) with $I_\parallel = -f''$ and $I_\perp = f'/d_{jk}$. Since the parameters for the magnetic bubble interactions are not known unlike the well-studied vortex interactions, we adopt the parameters for vortices: $\zeta = - 0.05 \, \omega_0$ and $\xi = - 0.15 \, \omega_0$, in which the minus sign represent the repulsive interactions. 

Equation~(\ref{eq:eom2}) is solved for three hexagonal lattices composed of distorted hexagons. Fig.~\ref{fig:fig3}(b) shows the one-dimensional dispersion for the coupled magnetic bubble gyration when the angles are $\theta_{jk} = \pi / 4, 7 \pi / 4, 3 \pi / 2$ for the sublattice sites $j$ at Y-shaped junctions. This case is similar to the vortex honeycomb lattice composed of regular hexagons, and thus exhibits the chiral edge mode rotating the boundary counterclockwise, same as the precession of individual magnetic bubbles. Fig.~\ref{fig:fig3}(d) shows the normal-mode dispersion when the angles are $\theta_{jk} = 0, \pi, 3 \pi / 2$ for the same sublattice sites $j$. In this case, the last term in Eq.~(\ref{eq:Haldane}) vanishes and thus the bulk band is gapless; the topological edge mode does not exist. Fig.~\ref{fig:fig3}(f) shows the dispersion when the angles are $\theta_{jk} = - \pi / 4, 7 \pi / 4, 3 \pi / 2$ for the aforementioned sublattice. In this case, the sign of the last term changes from the case of (a) and thereby exhibits the chiral edge mode rotating the boundary clockwise, opposite to the local precession of the constituent magnetic bubbles.

\emph{Discussion.}|We have shown that a honeycomb lattice of magnetic vortices and bubbles can exhibit a chiral edge mode via their coupled gyrations, the direction of which can be controlled by flipping the topological charge or by distorting the lattice geometry. The dispersions of the coupled vortex gyration have been investigated experimentally in several different arrangements including one-dimensional arrays of $5$ disks~\cite{HanSR2013} and two-dimensional arrays of $50 \times 50$ disks~\cite{BehnckePRB2015} by scanning transmission x-ray microscopy, which leads us to believe that experimental realization of our proposal for a vortex honeycomb lattice is within the current experimental reach. The experimental exploration of the chiral edge mode in a magnetic bubble lattice seems to be more challenging, as reflected in the relative lack of an experimental study on the dynamics of engineered magnetic bubble lattices.

We would like to mention that there is a class of ferrimagnets which allows us to thermally control the chirality of the edges modes. These are rare-earth transition-metal alloys such as GdFeCo and CoTb, possessing the special temperature referred to as the angular momentum compensation point, across which the gyromagnetic ratio changes its sign while keeping the magnetization finite~\cite{KirilyukRMP2010}. By changing the sign of the gyromagnetic ratio, we can flip the sign of the gyrotropic force in Eq.~(\ref{eq:eom})~\cite{KimAPL2017}. Therefore, when disks harboring vortices are made of such ferrimagnets, we should be able to change the chirality of the edge mode by varying the temperature across the compensation point, providing an example of temperature-driven topological phase transitions.

\begin{acknowledgments}
This work was supported by the Army Research Office under Contract No. W911NF-14-1-0016.
\end{acknowledgments}

\bibliography{/Users/evol/Dropbox/School/Research/master}

\end{document}